\pdfoutput=1

\documentclass[a4paper,12pt]{article}

\usepackage{url}
\usepackage{epsfig}
\usepackage{amsmath}
\usepackage{amssymb}
\usepackage{amsfonts}
\usepackage{bbm}
\usepackage[footnotesize]{caption}
\usepackage{graphicx}
\usepackage{mathrsfs}
\usepackage[font=scriptsize]{subcaption}
\usepackage{xcolor}
\usepackage{comment}
\usepackage{mathtools}
\usepackage{braket}
\usepackage{cite}
\usepackage{hyperref}
\usepackage{multirow}
\usepackage{relsize}
\usepackage{fullpage}
\usepackage{bbm}
\usepackage[T1]{fontenc}
\usepackage{pifont}
\usepackage{makecell}

\allowdisplaybreaks

\definecolor{darkpink}{RGB}{219, 48, 122}

\newcommand{\Dmq}{\Delta m^2}

\newcommand{\eVq}{\ensuremath{\text{eV}^2}}

\renewcommand{\Im}{\mathop{\mathrm{Im}}}
\usepackage{graphicx} 
\usepackage{mathtools}

\newcommand{\newc}{\newcommand}
\newc{\be}{\begin{equation}}
\newc{\ee}{\end{equation}}
\newc{\bea}{\begin{eqnarray}}
\newc{\eea}{\end{eqnarray}}
\newc{\simlt}{~\mbox{\smaller\(\lesssim\)}~}
\newc{\simgt}{~\mbox{\smaller\(\gtrsim\)}~}

\begin{document}

\begin{titlepage}

\begin{center}
{\bf\Large  
Orbifold Modular GUT of Flavour
} \\[12mm]
Francisco~J.~de~Anda$^{\ddagger}$%
\footnote{E-mail: \texttt{fran@tepaits.mx}},
Stephen~F.~King$^{\diamond}$%
\footnote{E-mail: \texttt{king@soton.ac.uk}}
\\[-2mm]

\end{center}
\vspace*{0.50cm}
\centerline{$^{\ddagger}$ \it
Tepatitl{\'a}n's Institute for Theoretical Studies, C.P. 47600, Jalisco, M{\'e}xico,}
\centerline{\it
Dual CP Institute of High Energy Physics, C.P. 28045, Colima, M\'exico.}
\vspace*{0.2cm}
\centerline{$^{\diamond}$ \it
School of Physics and Astronomy, University of Southampton,}
\centerline{\it
SO17 1BJ Southampton, United Kingdom.}
\vspace*{1.20cm}

\begin{abstract}
{\noindent  
We discuss an $SU(5)$ Grand Unified Theory (GUT) based on the 10d orbifold 
$(\mathbb{T}^2)^3/(\mathbb{Z}_4\times\mathbb{Z}_2)$ plus three modular $S_4$ groups with moduli at respective fixed points $i,i+2,\omega=e^{2i\pi/3}$. The resulting model has hierarchical quark and charged lepton mass matrices, arising from a double weighton mechanism, and reproduces the highly predictive Littlest Seesaw Mechanism in the neutrino sector.  
The down quark mass matrix has an upper triangular form, contributing to CKM mixing,
while the charged lepton mass matrix has a lower triangular form with suppressed contributions to PMNS mixing.
The orbifold yields successful $SU(5)$ breaking with doublet-triplet splitting of the Higgs multiplets.  }
\end{abstract}

\end{titlepage}

\section{Introduction}

The flavour puzzle of the Standard Model (SM) remains one of the most important questions in particle physics, both conceptually and pragmatically. From a fundamental point of view, the origin of hierarchical quark and charged lepton masses implies that the three families are somehow very different, even though their gauge interactions are identical. Practically this results in a plethora of unexplained parameters in the flavour sector of the SM, the more so following the discovery of neutrino mass and mixing, and surely too many for a fundamental theory. 
Recently there has been some progress in addressing the flavour problem, using modular symmetry using bottom-up approaches~\cite{Feruglio:2017spp} (for a recent review see~\cite{Ding:2023htn}). Although modular symmetry naturally arises from orbifold models, such constructions have been relatively neglected in the modular symmetry literature.

In a recent paper we developed a bottom-up approach to flavour models which combines modular symmetry with orbifold constructions~\cite{deAnda:2023udh}, especially on orbifolds in 10d which can provide three modular groups and three moduli fields in the low energy theory (below the compactification scales). Unlike 4d models, in such 10d orbifold models the values of the moduli are not completely free but are constrained geometry and symmetry. For example in the orbifold example $(\mathbb{T}^2)^3/(\mathbb{Z}_4\times\mathbb{Z}_2)$,
the fixed points, two of the moduli are constrained by the orbifold to be at $\tau =i$ (up to a discrete choice), while the third one is unconstrained by 
geometry but fixed by stabilisation arguments to be at $\tau =\omega =e^{2i\pi/3}$~\cite{King:2023snq}. 
The choice of the three moduli $\tau_1=i,\ \tau_2=i+2,\ \tau_3=\omega$ has recently been discussed in a 4d models of modular symmetry based on $S_4^3$~\cite{deMedeirosVarzielas:2022fbw,deMedeirosVarzielas:2023ujt} as well as 10d orbifold models also based on $S_4^3$~\cite{deAnda:2023udh}. This choice is of phenomenological interest since  
it leads to the highly predictive Littlest Modular Seesaw structure in the neutrino sector~\cite{Ding:2019gof,Ding:2021zbg}.  
\footnote{The Littlest Seesaw model is based on CSD($n\sim 3$)~\cite{King:2013iva,King:2015dvf,King:2016yvg,Ballett:2016yod,King:2018fqh,King:2013xba,King:2013hoa,Chen:2019oey},
and was recently reviewed in~\cite{Costa:2023bxw}.}

In the present paper we combine 10d orbifold Grand Unified Theories (GUTs) with modular symmetry with the goal of providing a unified theory of flavour.
Orbifold GUTs \cite{Hall:2001rz,Hebecker:2001ke,Hebecker:2002rc,Hebecker:2002re,Burrows:2009pi,Burrows:2010wz} have several advantages in terms of GUT symmetry breaking and simplified Higgs doublet-triplet splitting. It is therefore of interest to extend the above orbifold model of leptons~\cite{deAnda:2023udh}, to the case of orbifold GUTs, using the same orbifold and modular symmetry, in order to provide a complete theory including quark mass and mixing.
Here, then, we shall present an $SU(5)$ GUT based on a 10d orbifold $(\mathbb{T}^2)^3/(\mathbb{Z}_4\times\mathbb{Z}_2)$
with three modular $S_4$ symmetries with moduli at the fixed points $i,i+2,\omega$, which can successfully reproduce the Littlest Modular Seesaw
as in the model of leptons above.
The model also attempts to explain the quark and charged lepton mass hierarchies, using the weighton mechanism~\cite{King:2020qaj},
where the down quark mass matrix has an upper triangular form, contributing to CKM mixing, while the charged lepton mass matrix has a lower triangular form with suppressed contributions to PMNS mixing. This means that lepton mixing must originate entirely from the neutrino sector, thereby preserving the predictions of the Littlest Seesaw model. The same orbifold also yields successful $SU(5)$ breaking with doublet-triplet splitting of the Higgs multiplets.
The approach here follows an earlier work by two of us based on a 6d $SU(5)$ orbifold GUT with $A_4$ modular symmetry~\cite{deAnda:2018ecu} in which the resulting lepton sector had a $\mu - \tau$ symmetry.

The layout of the remainder of the paper is as follows. In section~\ref{2} we introduce the 10d orbifold of interest based on $(\mathbb{T}^2)^3/(\mathbb{Z}_4\times\mathbb{Z}_2)$, and discuss the resulting fixed points of interest, as well as the overlapping branes which permit interactions between the matter fields located there. Section~\ref{3} describes an $SU(5)$ model based on this orbifold, including GUT breaking, a discussion of fermion (matter) fields, Higgs doublet-triplet splitting, and the 10d and resulting 4d Lagrangians. We also discuss the resulting quark and lepton mass matrices and show that the model reproduces the Littlest seesaw model predictions. Section~\ref{4} concludes the paper.  In an Appendix we briefly review modular symmetry and modular forms at level $N=4$ corresponding to $S_4$.

\section{The orbifold $(\mathbb{T}^2)^3/(\mathbb{Z}_4\times\mathbb{Z}_2)$ }
\label{2}

We assume a 10d spacetime where the 6 extra dimensions are factorisable into 3 tori, each defined by one complex coordinate $z_i$ with $i=1,2,3,$ and compactified by the tori actions
\begin{equation}
 z_i\sim z_i+1,\ \ \ z_i\sim z_i+\tau_i,
 \label{eq:latac}
\end{equation}
where $\tau_i$ is a complex number to be fixed below.

The orbifold $(\mathbb{T}^2)^3/\mathbb{Z}_4\times\mathbb{Z}_2$, which is one from a restricted list of orbifolds which preserves supersymmetry~\cite{Fischer:2012qj},
as defined by the orbifolding actions 
\begin{equation}
\begin{split}
\theta_4&: (x,z_1,z_2,z_3)\sim (x,iz_1,-iz_2,z_3),\\
\theta_2&: (x,z_1,z_2,z_3)\sim (x,z_1,-z_2,-z_3).
\label{eq:orbac}
\end{split}
\end{equation}

Each of the orbifold operations $\theta_{2,4}$ can also be accompanied by a gauge transformation $P_{2,4}\in SU(5)$ respectively that satisfy
\begin{equation}
P_2^2=P_4^4=I,\ \ \ [P_2,P_4]=0.
\end{equation}

We assume that each tori has an independent discrete modular symmetry $S_4$.

The lattice of the extra dimensions are defined by the $\tau_i$. These are restricted geometrically by the orbifold action, as the orbifold action in Eq. \ref{eq:orbac} must be equivalent to some lattice translation as in Eq. \ref{eq:latac} i.e. its action over the lattice basis vectors $(1,\tau_i)$ must be a linear combination of the original lattice vectors, with integer coefficients.  
 Therefore there must exist integers $a_{1,2,3},b_{1,2,3},c_{1,2,3},d_{1,2,3}\in \mathbb{Z}$ such that
\begin{equation}
\begin{split}
(i,i\tau_{1,2})&=(a_{1,2}+b_{1,2}\tau_{1,2},c_{1,2}+d\tau_{1,2}),\\
(-1,-\tau_{3})&=(a_{3}+b_{3}\tau_{3},c_{3}+d\tau_{3}),\\
\end{split}
\label{1}
\end{equation}
In the present example, solving Eq.~\ref{1} gives,
\begin{equation}
\begin{split}
\tau_{1,2} &=i+n_{1,2},\ \ \ | \ \ \ n_{1,2}\in \mathbb{Z},\\
\tau_3 &\in \mathbb{C}.
\label{eq:tau24}
\end{split}
\end{equation}
this fixes $\tau_{1,2}=i+n_{1,2}$ where $n_{1,2}=0,1,2,3$, as we are assuming modular $S_4$ which constraints the available integers
\cite{deAnda:2023udh}. We choose $n_1=0$ and $n_2=1$. The $\tau_3$ can be any complex number.
We assume that there is a remnant $S_4$ symmetry on the 4d branes of the $z_3$ torus, which then fixes $\tau_3=\omega$ \cite{deAnda:2018ecu,deAnda:2018oik}, which is energetically favourable~\cite{King:2023snq}. 

This way all the moduli $\tau_i$ are fixed geometrically and the lattice is defined to be
\begin{equation}
 z_i\sim z_i+1,\ \ \ z_i\sim z_i+\tau_i,\ \ \ \tau_i=\{i,i+2,\omega\},
\end{equation}
where $\omega=e^{2i\pi/3}$.

Each orbifold action in Eq. \ref{eq:orbac}, leaves some invariant subspaces which are called fixed branes.

We want a minimal model where all fields can behave as modular forms (with different $\tau_i$ depending on their location) but can interact with each other, we will only use the 6d branes
\begin{equation}
\begin{split}
\mathbb{T}^2_A &=(x,z_1,0,0),\\
\mathbb{T}^2_B &=(x,0,z_2,0),\\
\mathbb{T}^2_C &=(x,0,0,z_3),
\label{eq:tbranes}
\end{split}
\end{equation}
where the $\mathbb{T}^2_A$ brane is left invariant by $\theta_2$, the $\mathbb{T}^2_B$ brane is left invariant by $\theta_2\theta_4^2$ and the $\mathbb{T}^2_C$ brane is left invariant by the action of $\theta_4^2$.
They all overlap with the origin brane, and this is where all interactions will happen.

\section{The model}
\label{3}

We build an $SU(5)$ model in 10d with $S_4$ modular symmetry. We assume the field content in Table \ref{ta:modelc}, where each field is located in different branes defined in Eq. \ref{eq:tbranes}.
\begin{table}[h]
\centering
\begin{footnotesize}
 \begin{tabular}{| l | c c c c c c c|c|}
\hline \hline
Field & $SU(5)$ & $S_4^A$ & $S_4^B$ & $S_4^C$ & \!$k_A$\! & \!$k_B$\! & \!$k_C$\! & Loc\\ 
\hline \hline
$F$ & $\textbf{5}$&$\mathbf{1}$ & $\mathbf{1}$ & $\mathbf{3}$ & 0 & 0 & $0$ &$\mathbb{T}^2_{C}$\\
$T_1$ & $\textbf{10}$& $\mathbf{1}$ & $\mathbf{1}$ & $\mathbf{1}$ & 0 & 0 & \!$1$\! &$\mathbb{T}^2_{C}$\\
$T_2$ & $\textbf{10}$& $\mathbf{1}$ & $\mathbf{1}$ & $\mathbf{1}$ & 0 & 0 & \!$1/2$\! &$\mathbb{T}^2_{C}$\\
$T_3$ & $\textbf{10}$&$\mathbf{1}$ & $\mathbf{1}$ & $\mathbf{1}$ & 0 & 0 & \!$0$\! &$\mathbb{T}^2_{C}$\\
$N_a^c$ & $\textbf{1}$&$\mathbf{1}$ & $\mathbf{1}$ & $\mathbf{1}$ & 0 & $-4$ & 0 &$\mathbb{T}^2_{B}$\\
$N_s^c$ & $\textbf{1}$& $\mathbf{1}$ & $\mathbf{1}$ & $\mathbf{1}$ & $-2$ & 0& 0 &$\mathbb{T}^2_{A}$\\
\hline 
$H_u$ & $\textbf{5}$& $\mathbf{1}$ & $\mathbf{1}$ & $\mathbf{1}$ & 0 & 0 & 0 & Bulk \\
$H_d$ & $\overline{\textbf{5}}$& $\mathbf{1}$ & $\mathbf{1}$ & $\mathbf{1}$ & 0 & 0 & $1/2$ & Bulk\\
$H_{45}$ & $\textbf{45}$& $\mathbf{1}$ & $\mathbf{1}$ & $\mathbf{1}$ & 0 & 0 & $1/2$ & Bulk \\
$H_{\overline{45}}$ & $\overline{\textbf{45}}$& $\mathbf{1}$ & $\mathbf{1}$ & $\mathbf{1}$ & 0 & 0 & $0$ & Bulk\\
$\Phi_{BC}$ & $\textbf{1}$& $\mathbf{1}$ & $\mathbf{3}$ & $\mathbf{3}$ & 0 & 0 & $0$& Bulk \\
$\Phi_{AC}$ & $\textbf{1}$&$\mathbf{3}$ & $\mathbf{1}$ & $\mathbf{3}$ & 0 & 0 & $0$ & Bulk \\
$\xi_F$ & $\textbf{1}$&$\mathbf{1}$ & $\mathbf{1}$ & $\mathbf{1}$ & 0 & 0 & \!$-5/2$\! &$\mathbb{T}^2_{C}$\\
$\xi_T$ & $\textbf{1}$&$\mathbf{1}$ & $\mathbf{1}$ & $\mathbf{1}$ & 0 & 0 & \!$-1/2$\! &$\mathbb{T}^2_{C}$\\

\hline \hline
\end{tabular}
\begin{tabular}{| l | c c c c c c|}
\hline \hline
Yuk/Mass &$S_4^A$ & $S_4^B$ & $S_4^C$ & \!$2k_A$\! & \!$2k_B$\! & \!$2k_C$\!\\
\hline \hline
$Y_e(\tau_3)$ & $\mathbf{1}$ & $\mathbf{1}$ & $\mathbf{3}$ & 0 & 0 & $6$ \\
$Y_\mu(\tau_3)$ & $\mathbf{1}$ & $\mathbf{1}$ & $\mathbf{3}$ & 0 & 0 & $4$ \\
$Y_\tau(\tau_3)$ & $\mathbf{1}$ & $\mathbf{1}$ & $\mathbf{3}$ & 0 & 0 & $2$ \\
$Y_a(\tau_2)$ & $\mathbf{1}$ & $\mathbf{3}$ & $\mathbf{1}$ & 0 & $4$ & 0  \\
$Y_s(\tau_1)$ & $\mathbf{3}$ & $\mathbf{1}$ & $\mathbf{1}$ & $2$ & 0 & 0\\\hline
$M_a(\tau_2)$ & $\mathbf{1}$ & $\mathbf{1}$ & $\mathbf{1}$ & 0 & $8$ & 0  \\
$M_s(\tau_1)$ & $\mathbf{1}$ & $\mathbf{1}$ & $\mathbf{1}$ & $4$ & 0 & 0 
\\
\hline \hline
\end{tabular}
\caption{Full list of the assumed fields of the model as well as they localization. The ones in the bulk are 10d chiral superfields while the ones in the defined branes are 6d chiral superfields. The modular forms in the second table are fixed by the representation and weights of the fields. The $H_{\overline{45}}$ is added to cancel anomalies and plays no other role in the low energy effective model.
}
\label{ta:modelc}
\end{footnotesize}
\end{table}

\subsection{GUT breaking}

In this model we will assume the $SU(5)$ breaking boundary conditions
\begin{equation}
P_2=I,\ \ \ P_4=diag(1,1,1,-1,-1).
\label{eq:p4}
\end{equation}

The 10d vector superfield fulfils the boundary condition
\begin{equation}
\begin{split}
\mathcal{V}(x,z_1,z_2,z_3)&=P_4 \mathcal{V}(x,iz_1,-iz_2,z_3)P_4,\\
\mathcal{V}(x,z_1,z_2,z_3)&=\mathcal{V}(x,z_1,-z_2,-z_3).
\label{eq:vp}
\end{split}
\end{equation}

As the 10d vector superfield decomposes into 4 4d superfields (1 vector and 3 left chiral superfields) $\mathcal{V}=\{V,\phi_{1,2,3}\}$ which fulfils the conditions \cite{Aranda:2020noz,Aranda:2020fkj,Aranda:2021eyn}
\begin{equation}
\begin{array}{ll}
V(x,z_1,z_2,z_3)=P_4 V(x,i z_1,-iz_2,z_3)P_4, & V(x,z_1,z_2,z_3)= V(x, z_1,-z_2,-z_3),\\
\phi_1(x,z_1,z_2,z_3)=iP_4 \phi_1(x,i z_1,-iz_2,z_3)P_4, & \phi_1(x,z_1,z_2,z_3)= \phi_1(x, z_1,-z_2,-z_3),\\
\phi_2(x,z_1,z_2,z_3)=-iP_4 \phi_2(x,i z_1,-iz_2,z_3)P_4, & \phi_2(x,z_1,z_2,z_3)= -\phi_2(x, z_1,-z_2,-z_3),\\
\phi_3(x,z_1,z_2,z_3)=P_4 \phi_3(x,i z_1,-iz_2,z_3)P_4, &\phi_3(x,z_1,z_2,z_3)= -\phi_3(x, z_1,-z_2,-z_3),
\end{array}
\end{equation}
where each 10d function decomposes into an infinite tower of KK modes. 
One can easily find the zero modes by finding the solutions for the prior equations when $z_1=z_2=z_3=0$. The only available zero modes are the SM gauge vector superfields.

\subsection{SM fermions}
All the SM fermions are located in the 6d brane $\mathbb{T}_C$ and therefore they are 6d chiral superfields which decompose as 2 4d chiral superfields (left and right) $F=\{F_L,F_R\}$. They comply with a single boundary condition (since $\theta_4$ doesn't act on the $\theta_4$ brane)
\begin{equation}
F_L(x,z_3)=F_L(x,-z_3),  F_R(x,z_3)= -F_R(x, -z_3),
\end{equation}
and therefore there is a single full left chiral superfield as a zero mode as desired.

\subsection{Higgs in the bulk and doublet-triplet splitting}
The Higgs is to be located in the bulk, it decomposse into 4 4d chiral superfields each fulfilling the boundary conditions (with a $-1$ charge under $\theta_4$)
\begin{equation}
\begin{array}{ll}
H_0(x,z_1,z_2,z_3)=-P_4 H_0(x,i z_1,-iz_2,z_3), & H_0(x,z_1,z_2,z_3)= H_0(x, z_1,-z_2,-z_3),\\
H_1(x,z_1,z_2,z_3)=-iP_4 H_1(x,i z_1,-iz_2,z_3), & H_1(x,z_1,z_2,z_3)= H_1(x, z_1,-z_2,-z_3),\\
H_2(x,z_1,z_2,z_3)=iP_4 H_2(x,i z_1,-iz_2,z_3), & H_2(x,z_1,z_2,z_3)= -H_2(x, z_1,-z_2,-z_3),\\
H_3(x,z_1,z_2,z_3)=-P_4 H_3(x,i z_1,-iz_2,z_3), &H_3(x,z_1,z_2,z_3)= -H_3(x, z_1,-z_2,-z_3),
\label{eq:htd}
\end{array}
\end{equation}
which works leaves only a single doublet as a zero mode, solving the doublet triplet splitting.

\subsection{Proton decay}

The GUT symmetry is broken by the orbifold boundary conditions defined in Eqs. \ref{eq:p4},\ref{eq:vp}, therefore GUT breaking happens at the compactification scale. As discussed in the previous sections, below compactification the model has the standard MSSM field content. This implies that gauge coupling unification happens at the standard scale of $M_{GUT}\sim 2\times 10^{16}\ {\rm GeV}$ \cite{Dimopoulos:1981yj}, which also defines the compactification scale.
\\
The GUT field content below compactification is the standard ones for a minimal SUSY $SU(5)$ GUT. Proton decay is therefore generated by the mediation of the extra gauge bosons $X,Y$ and the Higgs colour triplet $H_T$, however their couplings to the SM fields are affected by their extra dimensional profiles, changing the proton decay rate \cite{Hall:2001pg,Buchmuller:2004eg,Hebecker:2002rc}. In this model the $X,Y$ fields must fulfill the condition from Eq. \ref{eq:vp} 
\begin{equation}
X,Y(x,z_1,z_2,z_3)=-X,Y(x,iz_1,-iz_2,z_3),
\end{equation}
while the 4 Higgs colour triplets must fulfill the condition from Eq. \ref{eq:htd}
\begin{equation}
\begin{split}
H_{T0}(x,z_1,z_2,z_3)&=- H_{T0}(x,i z_1,-iz_2,z_3), \\
H_{T1}(x,z_1,z_2,z_3)&=-i H_{T1}(x,i z_1,-iz_2,z_3),\\
H_{T2}(x,z_1,z_2,z_3)&=i H_{T2}(x,i z_1,-iz_2,z_3), \\
H_{T3}(x,z_1,z_2,z_3)&=-H_{T3}(x,i z_1,-iz_2,z_3) .
\end{split}
\end{equation}
All of these fields have a KK tower whose lightest mass is $M_{GUT}$.
From these conditions it can be seen that all the proton decay mediating fields vanish at the $\mathbb{T}^2_C =(x,0,0,z_3)$ brane, where all the SM fermions are located. Therefore they don't couple directly to the SM. Their couplings to the SM fields can't be generated at loop level either,  as the available mediators, the SM gauge fields and the SM Higgs, do not match the necessary charges. 
\\
However there might be extra colour triplets inside the $H_{45}$. The representation decomposition $\mathcal{R}_{SU(5)}\to(\mathcal{R}_{SU(3)_C},\mathcal{R}_{SU(2)_L},q_{U(1)_Y})$ is
\begin{equation}
\textbf{45}\to (\textbf{1},\textbf{2},3)+(\textbf{3},\textbf{1},-2)+({\textbf{3}},\textbf{3},-2)+(\overline{\textbf{3}},\textbf{1},8)+(\overline{\textbf{3}},\textbf{2},-7)+(\overline{\textbf{6}},\textbf{1},-2)+({\textbf{8}},\textbf{2},3),
\end{equation}
with the normalization $q_{EM}=q_Y/6+T_3^{SU(2)}$. The orbifold boundary condition in Eq. \ref{eq:p4} can be written as 
\begin{equation}
P_4=diag(1,1,1,-1,-1)=diag(e^{i\pi 2},e^{i\pi 2},e^{i\pi 2},e^{-i\pi 3},e^{-i\pi 3})=e^{i\pi \hat{q}_Y},
\end{equation}
where $\hat{q}_Y$ is the hypercharge operator inside $SU(5)$. Just as the SM Higgs multiplet has a negative orbifold charge, the $H_{45}$ would transform 
\begin{equation}
H_{45\ 0}(x,z_1,z_2,z_3)=-e^{i\pi \hat{q}_Y} H_{45\ 0}(x,i z_1,-iz_2,z_3),
\end{equation}
so that the fields with odd $q_Y$ charge have zero modes (only the zeroth chiral superfield can have zero modes). Therefore the $T\sim (\overline{\textbf{3}},\textbf{2},-7)$ has a zero mode and may mediate proton decay. The model would also have its conjugate representation that these fields can have a renormalizable arbitrarily large mass term $\sim M_{45} H_{45}H_{\overline{45}}$. It is also suppressed by very small first family Yukawa couplings and these dimension 5 proton decay are further suppressed by SUSY breaking dependent terms \cite{Nath:2006ut,Bajc:2002bv,Emmanuel-Costa:2003szk,Babu:2010ej}. Therefore the proton decay process from this model can be well below the experimental constraints.
\\
The $H_{45}$ has been added to break the degeneracy of charged lepton and down quark masses. It can alternatively be broken by adding an adjoint ($\textbf{24}$) superfield that obtains a large VEV \cite{Bjorkeroth:2015ora}. If we assume this mechanism instead of the $H_{45}$ pair, the model would contain no mediators for proton decay.

\subsection{10d Lagrangian}

The 10d lagrangian must have dimension 
$[\mathcal{L}]=10,$ where the 10d scalars (like the one from $\Phi$ and $H$) have dimension $[\phi_{10}]=4$ while the 6d fermions (like any SM field) have dimension $[\psi_6]=5/2$.
Some interactions happen at a 6d brane and some at the origin 4d brane, which are indicated by a Dirac $\delta$ function. There are no renormalizable interactions and the superpotential will be written in terms of dimensionless coupling constants divided by a common scale $\Lambda$.

The VEV $\braket{\Phi}\sim\delta^i_j$ breaks two $S_4$ groups into the diagonal one \cite{deMedeirosVarzielas:2022fbw}, and its second power $\braket{\Phi}^2$ has non trivial structure. The relevant SM Yukawa 10d lagrangian that fill up the entire mass matrix structure of every fermion is
\begin{equation}
\begin{split}
\mathcal{L}_{10d} =&\mathcal{L}_{10d}^{(0)}+\mathcal{L}_{10d}^{(1)},\\ \\
\mathcal{L}_{10d}^{(0)}=&\left(\frac{y^{u}_{33}}{\Lambda^5}T_3T_3+\frac{y^{u}_{23}}{\Lambda^7}\xi_T T_2T_3\right)H_u\delta^6(z)\\
&+\left(\frac{y^{u}_{22}}{\Lambda^9}\xi^2_T T_2T_2+ \frac{y^{u}_{13}}{\Lambda^{9}}\xi_T^2 T_1 T_3+ \frac{y^{u}_{12}}{\Lambda^{11}}(\xi^3_T+\xi_F^3) T_1 T_2
\right)H_u\delta^6(z)\\
&+\left(\frac{Y_a}{\Lambda^9} F N_a^c\Phi_{BC}+\frac{Y_s}{\Lambda^9} F N_s^c\Phi_{AC}\right)H_u\delta^6(z)\\
&+ \left( \frac{Y_{5\tau}}{\Lambda^7}\xi_F F  T_3 + \frac{Y_{5\tau}^{'}}{\Lambda^9}\xi_F\xi_T F  T_2+ \frac{Y_{5\tau}^{''}}{\Lambda^{11}} \xi_F\xi_T^2 F  T_1 \right)H_{5d}\delta^6(z)\\
&+ \left( \frac{Y_{5\mu}}{\Lambda^9}\xi_F^2 F  T_2 + \frac{Y_{5\mu}^{'}}{\Lambda^{11}}\xi_F^2\xi_T F  T_1+ \frac{Y_{5e}}{\Lambda^{11}} \xi_F^3 F  T_1 \right)H_{5d}\delta^6(z)\\
&\quad + (H_{5d}^{(0)}\to H_{45d}^{(0)} \ {\rm terms})\\
& +\frac{M_a}{2}  N_a^c N_a^c  \delta^2(z_1)\delta^2(z_3) + \frac{M_s}{2}  N_s^c N_s^c \delta^2(z_2)\delta^2(z_3),\\ \\
\mathcal{L}_{10d}^{(1)}=&\left(\frac{y^{u}_{11}}{\Lambda^{13}}(\xi_T^4+\xi_F^3\xi_T) T_1 T_1+\frac{\tilde{y}^{u}_{33}}{\Lambda^{13}}T_3T_3\Phi^2_{AC,BC}\right)H_u\delta^6(z)\\
& +\left(\frac{\tilde{Y}_a}{\Lambda^{13}} F N_a^c\Phi_{BC}^2+\frac{\tilde{Y}_s}{\Lambda^{13}} F N_s^c\Phi_{AC}^2\right)H_u\delta^6(z)\\
&+\left(\frac{Y_a^{'}}{\Lambda^{13}} \xi_{T,X}^4F N_a^c\Phi_{BC}Y_\tau+\frac{Y_s^{'}}{\Lambda^{13}}\xi_{T,F}^4 F N_s^c\Phi_{AC}Y_\tau\right)H_u\delta^6(z)\\
&+\left(\frac{Y_a^{''}}{\Lambda^{21}} \xi_{T,X}^8 F N_a^c\Phi_{BC}Y_\mu+\frac{Y_s^{''}}{\Lambda^{21}}\xi_{T,F}^8 F N_s^c\Phi_{AC}Y_\mu\right)H_u\delta^6(z)\\
&+\left(\frac{Y_a^{''}}{\Lambda^{29}} \xi_{T,X}^{12}F N_a^c\Phi_{BC}Y_e+\frac{Y_s^{''}}{\Lambda^{29}}\xi_{T,F}^{12} F N_s^c\Phi_{AC}Y_e\right)H_u\delta^6(z)\\
&+ \left( \frac{Y_{5\mu}^{''}}{\Lambda^{15}}\xi_F\xi_{F,T}^4 F  T_3 + \frac{Y_{5e}^{'}}{\Lambda^{23}}\xi_F\xi_{F,T}^8 F  T_3+ \frac{Y_{5e}^{''}}{\Lambda^{17}} \xi_F^2\xi_{F,T}^4F  T_2 +\frac{\tilde{Y}_\tau}{\Lambda^{15}}\xi_F \Phi^2_{AC,BC}) F  T_3\right)H_{5}\delta^6(z)\\
&\quad + (H_{5d}^{(0)}\to H_{45d}^{(0)} \ {\rm terms}),\\
\end{split}
\end{equation}
where $\mathcal{L}_{10d}^{(1)}$ contains terms much more suppressed than $\mathcal{L}_{10d}^{(0)}$.

As is mentioned in the appendix \ref{app:modfor}, most singlet modular forms vanish. Therefore powers of $\xi$ are necessary to allow up quark masses and therefore they receive a non trivial structure.

We integrate the 6 extra dimensions and keep only the zero modes, where we will assume, for simplicity, a single radius for the 3 different tori $\sim R$. We will keep the same  When integrating the extra dimensions $SU(5)$ is broken but we will keep writing in terms of it for a simpler connection and add a superscript $(0)$ to denote that we are talking about zero modes.

Let us assume that the $\Phi$ fields do get a VEV $\braket{\Phi}$ above compactification but they do not have zero modes (they may have an $i$ charge under $\theta_4$), therefore it is a KK mode which gets a VEV.
\footnotesize{
\begin{equation}
\begin{split}
\mathcal{L}_{4d} =& \mathcal{L}_{4d}^{(0)}+\mathcal{L}_{4d}^{(1)},\\ \\
\mathcal{L}_{4d}^{(0)}=& \left(\left[\frac{y^{u}_{33}}{(2\pi\Lambda R)^5}\right]T_3^{(0)}T_3^{(0)}+\left[\frac{y^{u}_{23}}{(2\pi\Lambda R)^6}\right]\frac{(\xi_T^{(0)})}{\Lambda}T_2^{(0)}T_3^{(0)}\right.\\
&+\left.\left[\frac{y^{u}_{22}}{(2\pi\Lambda R)^7}\right]\frac{(\xi_T^{(0)})^2}{\Lambda^2}T_2^{(0)}T_2^{(0)}+\left[\frac{y^{u}_{13}}{(2\pi\Lambda R)^7}\right]\frac{(\xi_T^{(0)})^2}{\Lambda^2}T_1^{(0)}T_3^{(0)}+\left[\frac{y^{u}_{12}}{(2\pi\Lambda R)^8}\right]\frac{(\xi_{T,F}^{(0)})^3}{\Lambda^3}T_1^{(0)}T_2^{(0)}\right)H_u^{(0)}
\\
&+\left(\left[\frac{Y_a \braket{\Phi_{BC}}}{(2\pi \Lambda R)^5\Lambda^{4}}\right] F^{(0)} N_a^{c(0)} +\left[\frac{Y_s\braket{\Phi_{AC}}}{(2\pi \Lambda R)^5\Lambda^{4}}\right] F^{(0)} N_s^{c(0)} \right)H_u^{(0)}\\
&+ \left( \left[\frac{Y_{5\tau}}{(2\pi\Lambda R)^6}\right]\frac{(\xi_F^{(0)})}{\Lambda} F^{(0)}  T_3^{(0)} 
 +\left[\frac{Y_{5\tau}^{'}}{(2\pi\Lambda R)^7}\right] \frac{(\xi_F^{(0)})(\xi_T^{(0)})}{\Lambda^2} F^{(0)}  T_2^{(0)} +\left[\frac{Y_{5\tau}^{''}}{(2\pi\Lambda R)^8}\right]\frac{(\xi_F^{(0)})(\xi_T^{(0)})^2}{\Lambda^3} F^{(0)}  T_1^{(0)}
\right. \\
& +\left.\left[\frac{Y_{5\mu}}{(2\pi\Lambda R)^7}\right]\frac{(\xi_F^{(0)})^2}{\Lambda^2} F^{(0)}  T_2^{(0)} +\left[\frac{Y_{5\mu}^{'}}{(2\pi\Lambda R)^8}\right] \frac{(\xi_F^{(0)})^2(\xi_T^{(0)})}{\Lambda^3} F^{(0)}  T_1^{(0)} +\left[\frac{Y_{5e}}{(2\pi\Lambda R)^8}\right]\frac{(\xi_F^{(0)})^3}{\Lambda^3} F^{(0)}  T_1^{(0)} \right)H_{5d}^{(0)} \\
& + (H_{5d}^{(0)}\to H_{45d}^{(0)} \ {\rm terms})\\
&+ \frac{1}{2} M_a N_a^{c(0)} N_a^{c(0)} + \frac{1}{2} M_s N_s^{c(0)} N_s^{c(0)},
\\ \\
\mathcal{L}_{4d}^{(1)}=&
\left(\left[\frac{y^{u}_{11}}{(2\pi\Lambda R)^9}\right]\frac{(\xi_{F,T}^{(0)})^3(\xi_{T}^{(0)})}{\Lambda^4}T_1^{(0)}T_1^{(0)}+\left[\frac{\tilde{y}^{u}_{33}\braket{\Phi^2_{AC,BC}}^i_i}{(2\pi\Lambda R)^5\Lambda^{8}}\right]T_3^{(0)}T_3^{(0)}\right)H_u^{(0)}\\
& +\left(\left[\frac{\tilde{Y}_a^{}\braket{\Phi_{BC}^2}}{(2\pi\Lambda R)^5\Lambda^{8}}\right] F^{(0)} N_a^{c(0)} +\left[\frac{\tilde{Y}_s \braket{\Phi_{AC}^2}}{(2\pi \Lambda R)^5\Lambda^{8}}\right] F^{(0)} N_s^{c(0)}\right.\\
&+\left[\frac{Y_a^{'} \braket{\Phi_{BC}}Y_\tau}{(2\pi \Lambda R)^9\Lambda^{4}}\right] \frac{(\xi_{F,T}^{(0)})^4}{\Lambda^4}F^{(0)} N_a^{c(0)} +\left[\frac{Y_s^{'}\braket{\Phi_{AC}}Y_\tau}{(2\pi\Lambda R)^9\Lambda^{4}}\right] \frac{(\xi_{F,T}^{(0)})^4}{\Lambda^4}F^{(0)} N_s^{c(0)} \\
&+\left[\frac{Y_a^{''} \braket{\Phi_{BC}}Y_\mu}{(2\pi\Lambda R)^{13}\Lambda^{4}}\right] \frac{(\xi_{F,T}^{(0)})^8}{\Lambda^8}F^{(0)} N_a^{c(0)} +\left[\frac{Y_s^{''}\braket{\Phi_{AC}}Y_\mu}{(2\pi\Lambda R)^{13}\Lambda^{4}}\right] \frac{(\xi_{F,T}^{(0)})^8}{\Lambda^8}F^{(0)} N_s^{c(0)} \\
&\left.+\left[\frac{Y_a^{'''} \braket{\Phi_{BC}}Y_e}{(2\pi\Lambda R)^{17}\Lambda^{4}}\right] \frac{(\xi_{F,T}^{(0)})^8}{\Lambda^8}F^{(0)} N_a^{c(0)} +\left[\frac{Y_s^{'''}\braket{\Phi_{AC}}Y_e}{(2\pi\Lambda R)^{17}\Lambda^{4}}\right] \frac{(\xi_{F,T}^{(0)})^8}{\Lambda^8}F^{(0)} N_s^{c(0)} \right)H_u^{(0)}\\
& +\left(\left[\frac{Y_{5\mu}^{''}}{(2\pi\Lambda R)^{10}}\right]\frac{(\xi_F^{(0)})(\xi_{F,T}^{(0)})^4}{\Lambda^5} F^{(0)}  T_3^{(0)} +\left[\frac{Y_{5e}^{'}}{(2\pi\Lambda R)^{14}}\right] \frac{(\xi_F^{(0)})(\xi_{F,T}^{(0)})^8}{\Lambda^9}F^{(0)}  T_3^{(0)} \right.
\\
&\left. +\left[\frac{Y_{5e}^{''}}{(2\pi\Lambda R)^{11}}\right]\frac{(\xi_F^{(0)})^2(\xi_{F,T}^{(0)})^4}{\Lambda^6}F^{(0)}  T_2^{(0)}+\left[\frac{\tilde{Y}_{5\tau}^{''}\braket{\Phi^2_{AB,BC}}}{(2\pi \Lambda R)^6\Lambda ^{8}}\right]\frac{(\xi_F^{(0)})}{\Lambda} F^{(0)}  T_3^{(0)} \right)H_{5d}^{(0)}\ \ \ + (H_{5d}^{(0)}\to H_{45d}^{(0)} \ {\rm terms})
\end{split}
\end{equation}}
\normalsize
Where everything inside brackets is a dimensionless coupling constant or an effective modular form.

This lagrangian comes from a 10 dimensional one, which has implications on the dimensional factors appearing.
If this were a 4d lagrangian originally, we would have the replacements
\begin{equation}
\braket{\Phi}/\Lambda^4\to \braket{\Phi}/\Lambda,\ \ \ 2\pi R \Lambda \to 1.
\end{equation}
 Note that we assume that the $\varphi$ obtains its VEV above compactification. To enforce this, it does not have any zero modes and it is a KK mode which obtains a VEV \cite{Aranda:2019nac}. This implies it has dimensionality 4 and everytime it appears it is suppressed by $\Lambda^{-4}$, making it much more suppressed than the appearance of a 4d VEV like $\braket{\xi}$.

We assume that the dimensionless couplings reabsorb the factors $(2\pi\Lambda R)\approx 1$.
We define for simplicity
\begin{equation}
\tilde{\xi}=\frac{\braket{\xi^{(0)}}}{\Lambda},\ \ \ \tilde{\Phi}=\frac{\braket{\Phi}}{\Lambda^4}\sim\tilde{\xi}^4\ll 1.
\end{equation}
These terms fill up the entire mass matrices, however we will ignore terms of 
\begin{equation}
O(\Lambda^{-13})<\frac{m_u}{m_t}\sim 10^{-5},
\end{equation}
which will become approximate texture zeroes in the mass matrices. Therefore we ignore all the $\mathcal{L}^{(1)}$ terms. We can then simplify
\begin{equation}
\label{4dGUT}
\begin{split}
\mathcal{L}_{4d}^{(0)} =& \left(y^{u}_{33}T_3^{(0)}T_3^{(0)}+y^{u}_{23}\tilde{\xi}_T T_2^{(0)}T_3^{(0)}+y^{u}_{22}\tilde{\xi}_T^2T_2^{(0)}T_2^{(0)}+y^{u}_{13}\tilde{\xi}_T^2T_1^{(0)}T_3^{(0)}+y^{u}_{12}\tilde{\xi}_{T,F}^3 T_1^{(0)}T_2^{(0)}\right)H_u^{(0)}
\\
&+\left(Y_a \tilde{\Phi}_{BC}F^{(0)} N_a^{c(0)} +Y_s\tilde{\Phi}_{AC} F^{(0)} N_s^{c(0)} \right)H_u^{(0)}\\
&+ \left( Y_{5\tau}\tilde{\xi}_F F^{(0)}  T_3^{(0)} 
 +Y_{5\tau}^{'}\tilde{\xi}_F\tilde{\xi}_T F^{(0)}  T_2^{(0)} +Y_{5\tau}^{''}\tilde{\xi}_F\tilde{\xi}_T^2F^{(0)}  T_1^{(0)}
\right. \\
& \left.+Y_{5\mu} \tilde{\xi}_F^2 F^{(0)}  T_2^{(0)} +Y_{5\mu}^{'}\tilde{\xi}_F^2\tilde{\xi}_TF^{(0)}  T_1^{(0)} +Y_{5e}^{}\tilde{\xi}_F^3F^{(0)}  T_1^{(0)}
\right)H_{5d}^{(0)}\ \ \ + (H_{5d}^{(0)}\to H_{45d}^{(0)} \ {\rm terms})
\\
&+ \frac{1}{2} M_a N_a^{c(0)} N_a^{c(0)} + \frac{1}{2} M_s N_s^{c(0)} N_s^{c(0)} .
\end{split}
\end{equation}

\subsection{SM fermion mass matrices}

The structure of the modular forms \cite{Ding:2019gof} is described in the Appendix \ref{app:modfor}. Each modular form has a different representation and weight, as shown in Table \ref{ta:modelc}. Furthermore, each modular form comes from different tori so that they are defined by different moduli, giving each one different structure:
\begin{equation}
\begin{split}
Y_a&=Y_{\textbf{3}}^{(4)}(i+2)=y_a(0,1,-1)^T,\\
Y_s&=Y_{\textbf{3}}^{(2)}(i)=y_s(1,1+\sqrt{6},1-\sqrt{6})^T,\\
Y_\tau&=Y_{\textbf{3}}^{(2)}(\omega)=y_{\tau}(0,1,0)^T,\\
Y_\mu&=Y_{\textbf{3}}^{(4)}(\omega)=y_{\mu}(0,0,1)^T,\\
Y_e&=Y_{\textbf{3}I}^{(6)}(\omega)=y_e(1,0,0)^T.
\end{split}
\label{eq:modforms}
\end{equation}
we obtain the fermion masses. 
The $Y_{5e,5\mu,5\tau}$, $Y_{45e,45\mu,45\tau}$ and primed ones have the same modular form structure, which we simply write as
$Y_{e,\mu,\tau}$ respectively in Eq.\ref{eq:modforms}.
The lower case $y$'s are arbitrary complex dimensionless parameters. The primed, $5$ and $45$ subscript modular forms indicate different $y$ complex parameter but the same flavour structure.

The symmetric up-quark mass matrix originates from the $TTH_u$ couplings in Eq.~\ref{4dGUT},
\begin{equation}
M_u=\left(\begin{array}{ccc}
0 & y_{12}^u \tilde{\xi}_{T,F}^3 e^{i\phi_{u1}}&  y_{13}^u \tilde{\xi}_{T}^2\\
y_{12}^u \tilde{\xi}_{T,F}^3 e^{i\phi_{u1}}& y_{22}^u \tilde{\xi}_{T}^2 & y_{23}^u \tilde{\xi}_{T}e^{i\phi_{u2}}\\
 y_{13}^u \tilde{\xi}_{T}^2 &  y_{23}^u \tilde{\xi}_{T} e^{i\phi_{u2}}& y_{33}^u
\end{array}\right)v_u,
\end{equation}
where each $y$ is now an arbitrary real dimensionless constant. Phases can be redefined so that there are 5 real parameters and 2 phases.
This yields the approximate up-type quark mass hierarchies, $m_u\sim \tilde{\xi}_{T,F}^4v_u$, $m_c\sim \tilde{\xi}_{T}^2v_u$, $m_t\sim v_u$.

The $H_{45}$ breaks the charged lepton and down quark degeneracy. The $Y_{5e,5\mu,5\tau}$, $Y_{45e,45\mu,45\tau}$ and primed ones have the same modular form structure but different overall complex constants multiplying them. Therefore both are diagonal mass matrices but the actual masses are determined after Higgs mixing~\cite{Georgi:1979df}\footnote{The presence of both $H_{{45}}$ and $H_{\overline{45}}$ allows the mass term 
$M_{45}H_{{45}}H_{\overline{45}}$ which ensures that all the components of these Higgs fields are heavy, apart from the Higgs doublet component of $H_{\overline{45}}$ that mixes with the Higgs doublet contained in $H_{5d}$, to produce the light linear combination identified as the physical Higgs doublet $H_d$. In this way, the Higgs doublet-triplet splitting mechanism discussed earlier is sufficient to ensure one light physical combination 
of down-type Higgs doublets which we identify as $H_d$.},
\begin{equation}
\begin{array}{ll}
y_{e11}v_d=y_{5e} v_{d5}-3y_{5e}v_{d45}, & y_{d11}v_d=y_{5d} v_{d5}+y_{45d}v_{d45},\\
y_{e22}v_d=y_{5\mu} v_{d5}-3y_{5\mu}v_{d45}, & y_{d22}v_d=y_{5s} v_{d5}+y_{45s}v_{d45},\\
y_{e33}v_d=y_{5\tau} v_{d5}-3y_{5\tau}v_{d45}, & y_{d33}v_d=y_{5b} v_{d5}+y_{45b}v_{d45},\\
y_{e21}v_d=y'_{5\mu} v_{d5}-3y'_{5\mu}v_{d45}, & y_{d12}v_d=y'_{5s}v_{d5}+y'_{45s}v_{d45},\\
y_{e32}v_d=y'_{5\tau} v_{d5}-3y'_{5\tau}v_{d45}, & y_{d23}v_d=y'_{5b}v_{d5}+y'_{45b}v_{d45},\\
y_{e31}v_d=y''_{5\tau} v_{d5}-3y''_{5\tau}v_{d45}, & y_{d13}v_d=y''_{5b}v_{d5}+y''_{45b}v_{d45},\\
\end{array}
\end{equation}
where $v_d$ is an effective down Higgs VEV. 

The triangular down-quark and charged lepton mass matrices
originate from the $FTH_d$ couplings in Eq.~\ref{4dGUT},
\begin{equation}
M_d=\left(\begin{array}{ccc}
y_{d11}\tilde{\xi}_F^3 & y_{d12}\tilde{\xi}_F^2\tilde{\xi}_T & y_{d13}\tilde{\xi}_F\tilde{\xi}_T^2 \\
0 & y_{d22} \tilde{\xi}_F^2& y_{d23}\tilde{\xi}_F\tilde{\xi}_Te^{i\phi_{d2}}\\
0 & 0 & y_{d33}\tilde{\xi}_F
\end{array}\right)v_d,
\end{equation}
\begin{equation}
\label{Me}
M_e=\left(\begin{array}{ccc}
y_{e11} \tilde{\xi}_F^3& 0 & 0 \\
y_{e21} \tilde{\xi}_F^2 \tilde{\xi}_T & y_{e22}\tilde{\xi}_F^2 & 0\\
y_{e31} \tilde{\xi}_F\tilde{\xi}_T^2& y_{e32}\tilde{\xi}_F\tilde{\xi}_T e^{i\phi_{d1}}& y_{e33}\tilde{\xi}_F
\end{array}\right)v_d,
\end{equation}
where each matrix has 6 real parameters and 1 phase.
These yield the the approximate down-type quark and charged lepton mass hierarchies, 
$m_d\sim m_e\sim \tilde{\xi}_{F}^3v_d$, and $m_s\sim m_{\mu}\sim \tilde{\xi}_{F}^2v_d$, and $m_b\sim m_{\tau}\sim \tilde{\xi}_{F}v_d$.
 We have written each mass matrix in LR convention 
so that, upon diagonalisation, $M_d$ will yield left-handed mixing angles arising from the upper-right off-diagonal terms, while $M_e$ will yield approximately zero left-handed mixing angles (with non-zero right-handed mixing angles from the lower-left off-diagonal terms). This means that $M_d$ (as well as $M_u$) will both contribute approximately equally to the CKM mixing angles, 
while $M_e$ will not contribute appreciably to the PMNS mixing angles.

Finally the neutrino Dirac and Majorana mass matrices from $FNH_u$ and $N^cN^c$ terms are
\begin{equation}
M_D=\left(\begin{array}{cc}
0 & y_s\tilde{\Phi}_{AC}  \\
y_a\tilde{\Phi}_{BC} & y_s\tilde{\Phi}_{AC}(1-\sqrt{6}) \\
-y_a\tilde{\Phi}_{BC} & y_s\tilde{\Phi}_{AC}(1+\sqrt{6})
\end{array}\right)v_u,\ \ \ 
M_N=\left(\begin{array}{cc}
M_a & 0  \\
0 & M_s \end{array}\right),
\end{equation}
which has the structure of a type-I seesaw mechanism which generates effective mass matrix for the light neutrinos:
\begin{equation}
\footnotesize
m_\nu = M_D \cdot M_R^{-1} \cdot M_D^T = v_u^2 
\begingroup
\setlength\arraycolsep{15pt}
\begin{pmatrix}  
\dfrac{(y_s\tilde{\Phi}_{AC} )^2}{M_s} & \dfrac{(y_s\tilde{\Phi}_{AC} )^2 (2-n)}{M_s} &  \dfrac{(y_s\tilde{\Phi}_{AC} )^2n}{M_s} \\[12pt]
. & \dfrac{(y_a\tilde{\Phi}_{BC} )^2}{M_a} + \dfrac{(y_s\tilde{\Phi}_{AC} )^2 (2-n)^2}{M_s} & -\dfrac{(y_a\tilde{\Phi}_{BC} )^2}{M_a} + \dfrac{(y_s\tilde{\Phi}_{AC} )^2n(2-n)}{M_s} \\[12pt] 
. & . & \dfrac{(y_a\tilde{\Phi}_{BC} )^2}{M_a} + \dfrac{(y_s\tilde{\Phi}_{AC} )^2n^2}{M_s}
\end{pmatrix},
\endgroup
\label{eq:mnu_mee}
\end{equation}
where $n=1+\sqrt{6} \approx 3.45$. This can be redefined in terms of 3 independent physical parameters
\begin{equation}
\label{mnu}
m_\nu=m_a\left(\begin{matrix}
0 & 0& 0& \\
0 & 1 & -1\\
0 & -1 & 1
\end{matrix}\right)+m_be^{i\eta}\left(\begin{matrix}
1 & (2-n) & n& \\
(2-n) & (2-n)^2 & n(2-n)\\
n & n(2-n) & n^2
\end{matrix}\right),
\end{equation}
where
\begin{equation}
m_a=\left|\frac{v_u^2 (y_a\tilde{\Phi}_{BC} )^2}{M_a}\right|,\ \ \ m_b=\left|\frac{v_u^2 (y_s\tilde{\Phi}_{AC} )^2}{M_s}\right|
\end{equation}
which corresponds to flipped CSD($n$) with $n=1+\sqrt{6} \approx 3.45$ in the notation of ref.~\cite{Costa:2023bxw}.
Therefore the model has only these three parameters for the whole neutrino sector, where the PMNS mixing parameters do not receive any appreciable contribution from the charged lepton sector as mentioned above and discussed further below. This results in a highly predictive flipped CSD($1+\sqrt{6}$) setup~\cite{deAnda:2023udh,Costa:2023bxw} with an excellent fit to neutrino oscillation parameters involving three real input parameters to determine the three neutrino masses and the six parameters of the PMNS matrix, where one neutrino mass and one Majorana phase are predicted to be zero.

\begin{table}[h]
\centering
  \begin{footnotesize}
    \begin{tabular}{c|l|c|cc}
      \hline\hline
      && NuFit 5.2 $\pm 1\sigma$ & Natural Fit
      & Unrestricted fit
      \\
      \hline
      \rule{0pt}{4mm}\ignorespaces
      & $\theta_{12}/^\circ$
      & $33.41_{-0.72}^{+0.75}$ & 34.34
      & $34.32$ 
      \\[3mm]
      & $\theta_{23}/^\circ$
      & $49.1_{-1.3}^{+1.0}$ & 48.31
      & $48.95$ 
      \\[3mm]
      & $\theta_{13}/^\circ$
      & $8.54_{-0.12}^{+0.11}$ & 8.54
      & $8.54$ 
      \\[3mm]
      & $\delta/^\circ$
      & $197_{-25}^{+42}$ & 284
      & $253$
      \\[3mm]
      & $\dfrac{\Dmq_{21}}{10^{-5}~\eVq}$
      & $7.41_{-0.20}^{+0.21}$ & 7.42
      & $7.41$ 
      \\[3mm]
      & $\dfrac{\Dmq_{3\ell}}{10^{-3}~\eVq}$
      & $+2.511_{-0.021}^{+0.028}$ & 2.510
      & $2.511$ \\[3mm]
      \hline \hline &&&& \\[-3mm]
      & $\dfrac{m_a}{10^{-3}~{\rm eV}}$
      &  & $31.47$
      & $31.47$ \\[3mm]
      & $\dfrac{m_b}{10^{-3}~{\rm eV}}$
      &  
      &   2.32 & $-2.28$\\[3mm]
      & $\eta/\pi$
      & &  $1.24$ & $0.26$
       \\[3mm]\hline
      & $y_{e33}$
      & &  $0.46$
      &  $-0.06$\\[3mm]
      & $y_{e22}$
      & &  $0.13$
      & $0.89 $ \\[3mm]
      & $y_{e11}$
      & &  $0.003$
      &  $0.003$\\[3mm]
        & $y_{e32}$
      & &  $<0.7$
      &  $1.03$\\[3mm]
      & $y_{e31}$
      & &  $<0.7$
      & $0.04$ \\[3mm]
      & $y_{e21}$
      & &  $<0.7$
      &  $0.05$\\[3mm]
          & $\phi_{d1}/\pi$
      & &  $0-2$
      &  $1.33$\\[3mm]
      \hline
        & $\chi^2$
      & & $6.3$
      &  $1.56$\\
      \hline\hline
    \end{tabular}
  \end{footnotesize}
  \caption{Normal Ordering NuFit~5.2 values~\cite{Esteban:2020cvm} for the neutrino observables, and the best fit point from the model. 
  Note that the predictions for the PMNS mixing parameters for the natural fit are practically identical to that for the Littlest Seesaw Model in which the charged lepton mass matrix is diagonal.
  Assumed values for the fit: $\tan \beta=10$, $\tilde{\xi}_F=0.22$, and $\tilde{\xi}_T=0.41$. The charged lepton masses for both fits are: $m_e=0.486\ MeV$, $m_\mu=0.102\ GeV$, and $m_\tau=1.745$ which are at the center of the lepton masses run up to $M_{GUT}$ \cite{Antusch:2013jca}.}
  \label{ta:NuFit52}
\end{table}

One may ask how the fit changes due to the off-diagonal charged lepton mass matrix parameters in the lower-left of the mass matrix $M_e$ 
in Eq.~\ref{Me}. To address this question
we show two fits in Table \ref{ta:NuFit52}. First we assume that the off-diagonal terms in $M_e$ are smaller than the diagonal ones on the same row, which is a natural choice, since they are relatively suppressed by small expansion parameters. The results show that this generates quite a good for a wide range of off-diagonal terms, which supports the assertion that such terms are practically irrelevant. If we relax this condition, we can slightly improve the fit when they become larger than the diagonal ones, for instance by significantly increasing $y_{e32}$ relative to $y_{e33}$. We can see that even in this less natural case, the predictions emerging from the fit are similar to the first case. We conclude that the predictions for the PMNS parameters are quite independent of the parameters appearing in the charged lepton mass matrix. This means that the model maintains very similar predictions to the Littlest Modular Seesaw model where the charged lepton mass matrix is diagonal and neutrino masses and PMNS parameters are all determined from just the three input parameters in Eq.~\ref{mnu}. The present model in addition also accounts for the quark mass hierarchies and CKM mixing parameters, as well as the charged lepton mass hierarchies, in terms of two small expansion parameters arising from the weighton fields. However the charged fermion masses also depend on undetermined coefficients and so are not precisely predicted. Similarly the CKM angles, although generically predicted to be small, suppressed by powers of weighton expansion parameters, receive contributions from both the up and down quark sectors, and are not predicted or constrained. For this reason we do not show fits for the quark masses and CKM mixing parameters. A fit for all quark parameters are shown in Table \ref{ta:ckm}.

\begin{table}[h]
\centering
  \begin{footnotesize}
    \begin{tabular}{c|lc|lc}
      \hline\hline
      && Value $\pm 1\sigma$ & Parameter 
      \\
      \hline
      \rule{0pt}{4mm}\ignorespaces
      & $\theta^q_{12}/^\circ$
      & $13.04\pm 0.04$ & $y_{33}^u=0.519$ 
      \\[3mm]
      & $\theta^q_{23}/^\circ$
      & $2.24\pm 0.04$ & $y_{23}^u=-0.047$ 
      \\[3mm]
      & $\theta^q_{13}/^\circ$
      & $0.19\pm 0.01$ & $y_{13}^u=-0.007$ 
      \\[3mm]
      & $\delta^q/^\circ$
      & $68.75\pm 3.09$ & $y_{22}^u=-0.009$ 
      \\[3mm]
      & $m_u/MeV$
      & $0.50\pm 0.15$ & $y_{13}^u=-0.007$ 
      \\[3mm]
            & $m_d/MeV$
      & $0.84\pm 0.09$ &$y_{12}^u=0.001$ 
      \\[3mm]
            & $m_c/MeV$
      & $245.34\pm 8.58$ & $y_{33}^d=0.245$ 
      \\[3mm]
           & $m_s/MeV$
      & $16.70\pm 0.90$ & $y_{23}^d=-0.009$ 
      \\[3mm]
 & $m_b/GeV$
      & $0.939\pm 0.01$ & $y_{22}^d=0.019$  
      \\[3mm]
       & $m_t/GeV$
      & $90.48\pm 2.08$ & $y_{31}^d=-0.00003$ 
      \\[3mm]
             & 
      &  & $y_{21}^d=-0.009$  
      \\[3mm]
             & 
      &  & $y_{11}^d=-0.004$  
      \\[3mm]
             & 
      &  & $\phi_{u1}=0.88\pi$
      \\[3mm]
             & 
      &  & $\phi_{u1}=0.66\pi$
      \\[3mm]
             & 
      &  & $\phi_{d2}=1.12 \pi$
      \\[3mm]
      \hline\hline
    \end{tabular}
  \end{footnotesize}
  \caption{
  Assumed values for the fit: $\tan \beta=10$, $\tilde{\xi}_F=0.22$, and $\tilde{\xi}_T=0.41$. The quark masses and CKM parameters are evaluated at $M_{GUT}$ \cite{Antusch:2013jca}. The fit parameters generate the masses and CKM exactly at the center values with $\chi^2=0$. }
  \label{ta:ckm}
\end{table}

\section{Conclusion}
\label{4}

In this paper we have discussed 10d orbifold GUTS based on modular symmetry as candidates for a theory of Flavour.  To illustrate the approach, 
we presented an $SU(5)$ 10d orbifold theory, based on $S_4^3$ modular symmetry,
capable of giving quark and charged lepton mass hierarchies, using the weighton mechanism.  
The assumed $(\mathbb{T}^2)^3/(\mathbb{Z}_4\times\mathbb{Z}_2)$ orbifold yields three $S_4$ modular groups and three moduli at the fixed points $i,i+2,\omega=e^{2i\pi/3}$, which is a stable configuration. Using these fixed points, the model reproduces the highly predictive Littlest Seesaw Mechanism in the neutrino sector to very good approximation. The off-diagonal lower triangular entries in the charged lepton mass matrix yield only very small contributions to the left-handed PMNS angles, since the upper triangular entries are zero, a feature that we have verified numerically.
Although the model is very predictive in the neutrino and PMNS sectors, it does not make any predictions for CKM parameters.
The same orbifold also yields successful $SU(5)$ breaking with doublet-triplet splitting of the Higgs multiplets. 

\section*{Acknowledgements}
SFK acknowledges the STFC Consolidated Grant ST/L000296/1 and the European Union's Horizon 2020 Research and Innovation programme under Marie Sklodowska-Curie grant agreement HIDDeN European ITN project (H2020-MSCA-ITN-2019//860881-HIDDeN).

\appendix
\section{$S_4$ Modular forms}
\label{app:modfor}

With only two extra dimensions the single complex modulus $\tau$ has a modular symmetry $\overline{\Gamma}=SL(2,\mathbb{Z})$
The group $\overline{\Gamma}$ is the group of linear transformations which acts on the  modulus $\tau$ as follows,
\begin{equation}
\tau\rightarrow\gamma\tau=\frac{a\tau+b}{c\tau+d},~~\text{with}~~a, b, c, d\in\mathbb{Z},~~ad-bc=1,~~~\Im\tau>0\,.
\end{equation}
The modular group $\overline{\Gamma}$ can be generated by  $S$ and $T$
\begin{equation}
S:\tau\mapsto -\frac{1}{\tau},~~~~\quad T: \tau\mapsto\tau+1.
\end{equation}

From the infinite modular group the finite subgroup $\Gamma_N=PSL(2,\mathbb{Z})/\Gamma (N)$ may be obtained.
A crucial element of the modular invariance is the modular form $f(\tau)$ of weight $2k$ and level $N$. The modular form $f(\tau)$ is a holomorphic function of the modulus $\tau$ and it is required to transform under the action of $\overline{\Gamma}(N)$ as,
\begin{equation}
f\left(\frac{a\tau+b}{c\tau+d}\right)=(c\tau+d)^{2k}f(\tau)
\end{equation}
The modular forms of level $N=4$ have been constructed in~\cite{Penedo:2018nmg,Novichkov:2018ovf} .

The associated finite modular group $\Gamma_4$ with generators $S$ and $T$ which fulfill the following  rations
\begin{equation}
S^2=(ST)^3=(TS)^3=T^4=1\,.
\end{equation}

The modular forms are built from the functions $Y_{i}(\tau)$ with $i=1,...,5$ \cite{Novichkov:2018ovf}.
The ones we use are:
\begin{equation}
\begin{split}
Y_\textbf{3}^{(2)}=&\left(\begin{array}{c} Y_3\\Y_4\\ Y_5
\end{array}\right),\\
Y_\textbf{3}^{(4)}=&\left(\begin{array}{c} Y_1 Y_4+Y_2Y_5\\Y_2 Y_3+ Y_1 Y_5\\ Y_1Y_3+Y_2 Y_4
\end{array}\right),\\
Y_{\textbf{3}I}^{(6)}=&\left(\begin{array}{c} 2Y_1Y_2Y_3\\2Y_1Y_2Y_4\\ 2Y_1Y_2Y_5
\end{array}\right),\\  Y_{\textbf{3}II}^{(6)}=&\left(\begin{array}{c} Y_2^2Y_4+Y_1^2Y_5\\Y_1^2Y_3+Y_2^2Y_5\\ Y_2^2Y_3+Y_1^2Y_4
\end{array}\right),\\
Y_\textbf{1}^{(0)}=& 1   ,\\
Y_\textbf{1}^{(2)}=&0    ,\\
Y_\textbf{1}^{(4)}=& 2Y_1Y_2   ,\\
Y_\textbf{1}^{(6)}=&Y_1^3+Y_2^3 ,\\
Y_\textbf{1}^{(8)}=& Y_1^2 Y_2^2 ,\\
Y_\textbf{1}^{(10)}=&  Y_2 Y_1^4+Y_2^4 Y_1  ,\\
\end{split}
\end{equation}

At $\tau=\omega$ we have
\begin{equation}
Y_1(\omega)=Y_3(\omega)=Y_5(\omega)=0,\ \ \ Y_2(\omega)=\alpha,\ \ \ Y_4(\omega)=\alpha(1-\omega^2),
\end{equation}
where $\alpha\approx 1.79288$.

At $\tau=i$ we have
\begin{equation}
Y_1(i)=2\beta,\ \ \ Y_2(i)=1.73\beta,\ \ \ Y_3(i)=\beta,\ \ \ Y_4(i)=(1+\sqrt{6})\beta,\ \ \ Y_5(i)=(1-\sqrt{6})\beta,\ \ \
\end{equation}
where $\beta\approx 0.696$.
\\
At $\tau=i+2$ we have
\begin{equation}
Y_1(i+2)\approx 3.115,\ \ \ Y_2(i+2)\approx 1.206,\ \ \ Y_3(i+2)\approx 10.449,\ \ \ Y_4(i+2)\approx 2.698,\ \ \ Y_5(i+2)\approx 1.206.\ \ \
\end{equation}

\end{document}